# Magnetar as the Central Engine of AT2018cow: Optical, Soft X-Ray, and Hard X-Ray Emission

Long Li[1,2], Shu-Qing Zhong[1,2], Di Xiao[3], Zi-Gao Dai[1,2], Shi-Feng Huang[1,2], and Zhen-Feng Sheng[4]
[1] Department of Astronomy, University of Science and Technology of China, Hefei 230026, People's Republic of China; lilong1125@ustc.edu.cn, daizg@ustc.edu.cn
[2] School of Astronomy and Space Science, University of Science and Technology of China, Hefei 230026, People's Republic of China
[3] Purple Mountain Observatory, Chinese Academy of Sciences, Nanjing 210023, People's Republic of China
[4] Institute of Deep Space Sciences, Deep Space Exploration Laboratory, Hefei 230026, People's Republic of China

## Abstract

AT2018cow is the most extensively observed and widely studied fast blue optical transient to date; its unique observational properties challenge all existing standard models. In this paper, we model the luminosity evolution of the optical, soft X-ray, and hard X-ray emission, as well as the X-ray spectrum of AT2018cow with a magnetar-centered engine model. We consider a two-zone model with a striped magnetar wind in the interior and an expanding ejecta outside. The soft and hard X-ray emission of AT2018cow can be explained by the leakage of high-energy photons produced by internal gradual magnetic dissipation in the striped magnetar wind, while the luminous thermal UV/optical emission results from the thermalization of the ejecta by the captured photons. The two-component energy spectrum yielded by our model with a quasi-thermal component from the optically thick region of the wind superimposed on an optically thin synchrotron component well reproduces the X-ray spectral shape of AT2018cow. The Markov Chain Monte Carlo fitting results suggest that in order to explain the very short rise time to peak of the thermal optical emission, a low ejecta mass $M_{\rm ej} \approx 0.1\,M_\odot$ and high ejecta velocity $v_{\rm SN} \approx 0.17c$ are required. A millisecond magnetar with $P_0 \approx 3.7$ ms and $B_p \approx 2.4 \times 10^{14}$ G is needed to serve as the central engine of AT2018cow.

*Unified Astronomy Thesaurus concepts:* Magnetars (992); X-ray transient sources (1852); Supernovae (1668); Core-collapse supernovae (304)

## 1. Introduction

High-cadence surveys over the last decade have discovered a plethora of rapidly evolving transients suggested to have diverse physical origins (see Inserra 2019, for a review). "Fast blue optical transients" (FBOTs; Drout et al. 2014; Arcavi et al. 2016; Tanaka et al. 2016; Pursiainen et al. 2018; Ho et al. 2023) are a subset of those typically characterized by a blue color ($g - r \lesssim -0.2$ mag at peak) and a rapid light curve (time above half-maximum luminosity $t_{1/2} \lesssim 12$ days). AT2018cow (Prentice et al. 2018; Margutti et al. 2019; Perley et al. 2019), the most widely studied FBOT, exhibits peculiar observational properties that challenge all standard models.

AT2018cow boasts the richest set of observations among FBOTs, with emissions ranging from radio to hard X-rays (Prentice et al. 2018; Rivera Sandoval et al. 2018; Ho et al. 2019; Kuin et al. 2019; Margutti et al. 2019; Perley et al. 2019; Nayana & Chandra 2021). The UV+$UBV$ emission from AT2018cow can be aptly modeled using blackbody radiation spectrum, with no evidence of a transition to the nebular phase within 90 days (Margutti et al. 2019). However, the near-infrared+$RI$ emission clearly exceeds the thermal blackbody emission, suggesting that they might stem from a different underlying mechanism or source (Margutti et al. 2019; Perley et al. 2019; Metzger & Perley 2023). Within a few days the bolometric luminosity rises to a peak of $L_{\rm bol} \approx 4 \times 10^{44}$ erg s$^{-1}$ and then declines with $L_{\rm bol} \propto t^{-2.5}$. AT2018cow also exhibited a luminous and highly variable soft X-ray emission in the 0.3–10 keV range, with a luminosity measured at $L_X \sim 10^{43}$ erg s$^{-1}$ initially. The soft X-ray flaring shows short variability timescales of a few days, overlaying an initial gradual decay $\propto t^{-1}$ that later steepens to a faster $\propto t^{-4}$ rate (Rivera Sandoval et al. 2018; Kuin et al. 2019; Margutti et al. 2019). Furthermore, a hard X-ray component of emission, spanning 20–200 keV with distinct temporal and spectral properties, was detected at ∼8 days and dissipated by ∼17 days (Margutti et al. 2019). The spectra of joint soft and hard X-rays can be described by an absorbed power-law component superposed by an additional absorbed cutoff power-law component (Margutti et al. 2019). The total luminosity from the thermal UV/optical and X-ray emissions decays according to $\propto t^{-2}$ (Margutti et al. 2019). In addition, AT2018cow shows bright radio and millimeter emission, consistent with self-absorbed synchrotron radiation from shock interaction between fast ejecta and dense external medium (Ho et al. 2019; Margutti et al. 2019; Nayana & Chandra 2021).

The optical emission of AT2018cow cannot be powered dominantly by the radioactive decay of $^{56}$Ni, which is the standard energy source model for normal supernovae (SNe), as its large bolometric peak luminosity and short rise to peak time require a $^{56}$Ni mass of ∼$6\,M_\odot$ but an ejecta mass of $\lesssim 1\,M_\odot$. There are so many theories and explanations in the literature about the progenitor system and energy source of AT2018cow, including a successful SN with a low-mass ejecta giving birth to a millisecond magnetar as the central engine (Fang et al. 2019; Margutti et al. 2019; Mohan et al. 2020); accretion-induced collapse (AIC) of the product of double white dwarf (WD) merger between ONeMg WD and another WD giving birth to a millisecond magnetar (Lyutikov & Toonen 2019; Lyutikov 2022); AIC of a WD in a binary with a nondegenerate companion giving birth to a millisecond magnetar (Yu et al. 2019); an ejecta–circumstellar material (CSM) interaction similar to that of Type Ibn/IIn SNe (Fox & Smith 2019); a shocked jet in a core-collapse SN (Gottlieb et al. 2022); a pulsational pair-instability SN (Leung et al. 2020); a failed SN produces mass ejection through the accretion disk wind of a black hole (BH; Margutti et al. 2019; Perley et al. 2019; Quataert et al. 2019; Piro & Lu 2020; Uno & Maeda 2020; Antoni & Quataert 2022); a wind-reprocessed transient (Piro & Lu 2020; Uno & Maeda 2020); a tidal disruption event (TDE) of a star by an intermediate-mass BH (Perley et al. 2019); a TDE of a white dwarf by an intermediate-mass BH (Kuin et al. 2019); a TDE of a star by a stellar-mass BH (Kremer et al. 2021); a common envelope jets SN (Soker et al. 2019; Soker 2022); an ejecta–CSM interaction driven by the prompt merger of a BH–neutron star (NS) system after a common envelope phase (Schrøder et al. 2020); a delayed merger of a BH/NS system after a common envelope interaction (Metzger 2022).

If a millisecond magnetar serves as the central engine of AT2018cow, the thermal UV/optical emission is likely to be powered by the absorption and reprocessing of high-energy particles and radiation from the central magnetar by the ejecta. Instead, the observed X-ray emission may be contributed by radiation from a nebula of relativistic electron–positron pairs and radiation behind the expanding ejecta (Kotera et al. 2013; Metzger et al. 2014; Murase et al. 2015), or directly by radiation resulting from internal dissipation in the magnetar wind (e.g., Beniamini & Giannios 2017) that escapes from the ejecta without being thermalized. Vurm & Metzger (2021) calculated magnetar-driven SN light curves and energy spectra based on radiative transfer simulations that follow the thermalization and escape of high-energy radiation from the nebula. Although their model may be applicable to FBOTs, it has not been applied to AT2018cow yet. For the latter case, no literature has yet examined this scenario in detail in the context of SNe or FBOTs.

A newborn magnetar can generate a powerful Poynting flux–dominated wind, and as it propagates outward, this wind can be accelerated with the dissipation of its internal energy through mechanisms such as reconnection, eventually forming an ultrarelativistic wind dominated by electron–positron pairs at large radii (e.g., Coroniti 1990). In the field of gamma-ray bursts (GRBs), such a powerful magnetar wind can even reshape the GRB afterglow through energy injection or termination shock (e.g., Dai & Lu 1998a, 1998b; Zhang & Mészáros 2001; Dai 2004; Yu & Dai 2007; Geng et al. 2018; Li & Dai 2021). The detailed acceleration and dissipation processes of the Poynting flux–dominated outflow are affected by magnetic field configuration. In one possible configuration, the outflow may be composed of wound-up magnetic field lines in a "striped wind" geometry. The magnetic energy is gradually dissipated through reconnection, consequently accelerating the electrons and producing radiation (Spruit et al. 2001; Drenkhahn 2002; Drenkhahn & Spruit 2002; Giannios & Spruit 2005; Giannios 2006, 2008, 2012; Metzger et al. 2011; Beniamini & Piran 2014; Sironi & Spitkovsky 2014; Kagan et al. 2015; Sironi et al. 2015). Recent studies have shown that such radiation is significant in the X-ray/gamma-ray bands (Beniamini & Giannios 2017; Xiao & Dai 2017, 2019; Xiao et al. 2018, 2019). Moreover, the spectrum of the radiation consists of a quasi-thermal spectrum with a peak at tens to hundreds of keV superimposed on a synchrotron spectrum, which is quite similar to the case of AT2018cow. This motivates us to apply such an internal dissipation process to the context of SNe/FBOTs to explain the multiwavelength observations of AT2018cow.

This paper is organized as follows. In Section 2, we present our semianalytic model. In Section 3, we fit the multi-wavelength light curves and spectrum of AT2018cow using the semianalytic model and the Markov Chain Monte Carlo (MCMC) approach, and present the fitting results. Our discussion and conclusions can be found in Section 4.

## 2. The Model

In this study, we consider a system composed of two separate regions: the inner magnetar wind and the external ejecta. We assume that the spin and magnetic axes are orthogonal, which is plausible for a newborn magnetar (e.g., Lander & Jones 2020). In this situation, the dissipation region of the outflow is quasi-spherical. For the ejecta, several observations indicate that AT2018cow exhibits at least two distinct components: high-velocity ($\sim 0.1c$), low-density ejecta lacking hydrogen and helium along the polar direction, and low-velocity ($\sim 0.02c$) dense ejecta enriched in hydrogen and helium in the equatorial plane (e.g., Margutti et al. 2019). However, for simplicity and to maintain a focus on the primary phenomena we aim to explore, our model currently considers only a single component, assuming a spherically symmetric ejecta structure. Furthermore, we ignore any radiation from the nebula, and the absorption and reprocessing of radiation from the magnetar wind by the nebula. Further studies and iterations of the model can potentially incorporate the complexity of multiple ejecta components as well as the treatment of nebula to provide a more nuanced understanding of AT2018cow.

### 2.1. Emission from the Striped Magnetar Wind

A newborn millisecond magnetar can produce a Poynting flux–dominated outflow carrying a globally ordered magnetic field. When the magnetic field axis is misaligned with the spin axis, the magnetic field configuration of the outflow has a "striped wind" geometry (Coroniti 1990; Spruit et al. 2001). As the flow propagates outward, magnetic energy dissipates due to the reconnection of field lines with opposite orientations. Such magnetic dissipation leads to two effects: first, it causes the magnetic pressure to decrease outward so that the outflow is accelerated and the Poynting flux is converted to bulk kinetic energy (Spruit et al. 2001; Drenkhahn 2002; Drenkhahn & Spruit 2002; Giannios & Spruit 2005); second, it can lead to radiation directly as electrons in the outflow are accelerated due to reconnection (Giannios & Spruit 2005; Beniamini & Piran 2014; Sironi & Spitkovsky 2014; Kagan et al. 2015). Assuming that the total luminosity of the outflow is characterized by the spin-down luminosity dominated by magnetic dipole radiation, the evolution with time can be written as (Shapiro & Teukolsky 1983)

$$L_{\rm sd} = L_0 \left(1 + \frac{t}{\tau_{\rm sd}}\right)^{-2}, \quad (1)$$

where

$$L_0 = \frac{B_p^2 R^6 \Omega_0^4}{6c^3} \simeq 9.6 \times 10^{48} B_{p,15}^2 P_{0,-3}^{-4} R_6^6 \,\, {\rm erg \,\, s^{-1}} \quad (2)$$

is the initial spin-down luminosity and

$$\tau_{\rm sd} = \frac{3c^3 I}{B_p^2 R^6 \Omega_0^2} \simeq 2.1 \times 10^3 I_{45} B_{p,15}^{-2} P_{0,-3}^2 R_6^{-6} \text{ s} \tag{3}$$

is the spin-down timescale. Here $\Omega_0 = 2\pi/P_0$ and $P_0$ are the initial angular velocity and initial spin period of the magnetar, $B_p$ is the surface magnetic field strength at the polar cap region, and $I$ and $R$ are the moment of inertia and radius, respectively. In order to calculate the radiation at any radius, we need to know the variation of the outflow parameters with radius. At a given radius, the Poynting flux luminosity per steradian can be written as (Giannios & Spruit 2005; Beniamini & Giannios 2017)

$$L_B = c \frac{(rB)^2}{4\pi} = \frac{L_{\rm sd}}{4\pi} \left[ 1 - \frac{\Gamma}{\Gamma_{\rm sat}} \right], \tag{4}$$

where $B$ and $\Gamma$ are the magnetic field strength and bulk Lorentz factor of the outflow at radius $r$, respectively. $\Gamma_{\rm sat}$ is the bulk Lorentz factor at the saturation radius $r_{\rm sat} = \lambda \Gamma_{\rm sat}^2/(6\epsilon) = 1.7 \times 10^{13} \Gamma_{\rm sat,3}^2 (\lambda/\epsilon)_8$ cm (Beniamini & Giannios 2017), where $\lambda \sim cP = 3 \times 10^7 P_{-3}$ cm is the characteristic width of the layer of alternating magnetic polarity in the striped wind (Coroniti 1990; Spruit et al. 2001; Drenkhahn 2002; Drenkhahn & Spruit 2002), and $\epsilon \sim (0.1\text{--}0.25)c$ is the velocity of plasma flow to the reconnection layer (Lyubarsky 2005; Guo et al. 2015; Liu et al. 2015). The dissipation below the photospheric radius is reprocessed into a quasi-thermal emission. The thermal luminosity decreases as $L_{\rm th}(r) \propto r^{-4/9}$ as the comoving temperature decreases as $T' \propto r^{-7/9}$ (Giannios & Spruit 2005). By substituting the energy dissipation rate $d\dot{E} = -(dL_B/dr)dr$, the thermal luminosity can be written as (Giannios & Spruit 2005; Beniamini & Giannios 2017)

$$L_{\rm ph} = \int_0^{r_{\rm ph}} \frac{1}{2} \left( \frac{r}{r_{\rm ph}} \right)^{4/9} d\dot{E}$$
$$= 1.7 \times 10^{47} L_{\rm sd,49}^{6/5} \Gamma_{\rm sat,3}^{-1} \left( \frac{\lambda}{\epsilon} \right)_8^{-1/5} \text{ erg s}^{-1} \text{ sr}^{-1},$$

and the corresponding temperature (Xiao & Dai 2017)

$$T_{\rm ph} = 42 L_{\rm sd,49}^{1/10} \Gamma_{\rm sat,3}^{1/4} \left( \frac{\lambda}{\epsilon} \right)_8^{-7/20} \text{ keV}. \tag{5}$$

Here $r_{\rm ph} = 7.3 \times 10^9 L_{\rm sd,49}^{3/5} \Gamma_{\rm sat,3}^{-1} (\lambda/\epsilon)_8^{2/5}$ cm is the photospheric radius derived by setting the Thomson electron scattering depth $\tau(r_{\rm ph}) = 1$ (Beniamini & Giannios 2017).

Above the photospheric radius, the emission is dominated by synchrotron radiation. Particle-in-cell simulations show that magnetic reconnection accelerates electrons to a power-law distribution of the particle energy with an energy spectral index $p$ (Sironi & Spitkovsky 2014; Guo et al. 2015; Kagan et al. 2015; Werner et al. 2016). The minimum Lorentz factor of electrons written as ($p > 2$; Beniamini & Giannios 2017)

$$\gamma_m = \frac{p-2}{p-1} \frac{\epsilon_e}{2\xi} \sigma(r) \frac{m_p}{m_e}, \tag{6}$$

where $\epsilon_e$ is the fraction of the dissipated energy per electron, and $\xi \simeq 0.2$ is the fraction of the electrons accelerated in the reconnection sites (Sironi et al. 2015). The maximum Lorentz factor derived by setting the acceleration timescale $t_{\rm acc} = (\gamma_e m_e c^2)/(q\epsilon B'c)$ due to reconnection (Giannios 2010) equal to the cooling timescale $t_{\rm syn} = (6\pi m_e c)/(\sigma_T B'^2 \gamma_e)$ of the synchrotron radiation can be written as

$$\gamma_{\rm max} = \left( \frac{6\pi q \epsilon}{\sigma_T B'} \right)^{1/2}, \tag{7}$$

where $\sigma_T$ is the Thomson scattering cross section, $q$ is the electron charge, and $B'$ is the comoving magnetic field strength of the outflow. The synchrotron radiation spectrum undergoing synchrotron cooling can be expressed in the form a multi-segment broken power law. Initially, the electrons are in the fast cooling regime and the radiation spectrum is written as (Sari et al. 1998)

$$L_\nu^{\rm syn} = L_{\nu,\rm max}^{\rm syn} \begin{cases} (\nu/\nu_c)^{1/3} & \nu < \nu_c = \frac{1}{1+z} \frac{72\pi e m_e c^3 \Gamma^3}{\sigma_T^2 B'^3 r^2}, \\ (\nu/\nu_c)^{-1/2} & \nu_c < \nu < \nu_m = \frac{1}{1+z} \Gamma \gamma_m^2 \frac{qB'}{2\pi m_e c}, \\ (\nu_m/\nu_c)^{-1/2} (\nu/\nu_m)^{-p/2} & \nu_m < \nu < \nu_{\rm max} = \frac{1}{1+z} \Gamma \gamma_{\rm max}^2 \frac{qB'}{2\pi m_e c}, \end{cases} \tag{8}$$

where

$$L_{\nu,\rm max}^{\rm syn} = (1+z) \frac{m_e c^2 \sigma_T \Gamma B' N_e(r)}{3q}. \tag{9}$$

Here $N_e(r)$ is the number of electrons at radius $r$ (per steradian). The synchrotron flux is self-absorbed below the frequency $\nu_a$. The self-absorption frequency $\nu_a$ and corresponding electron Lorentz factor $\gamma_a$ satisfy

$$\frac{2\nu_a^2}{c^2} \gamma_a \Gamma m_e c^2 \frac{\pi r^2}{\Gamma^2} = \frac{L_{\nu_a}^{\rm syn}}{(1+z)^3}, \tag{10}$$

where

$$\nu_a = \frac{1}{1+z} \Gamma \gamma_a^2 \frac{qB'}{2\pi m_e c}. \tag{11}$$

At $r_{\rm ph}$ usually $\nu_a > \nu_c$, and then $\nu_a$ decreases to below $\nu_c$ at a certain radius $r_{\rm cr}$. We also define a transition radius $r_{\rm tr}$ from fast to slow cooling by making $\nu_m = \nu_c$. Since the spectral slope below $\nu_a$ is $L_\nu \propto \nu^{11/8}$ (Granot & Sari 2002), the entire synchrotron radiation spectrum can be written in the following. For $r_{\rm ph} < r \leqslant r_{\rm cr}$,

$$L_\nu^{\rm syn} = L_{\nu_a}^{\rm syn} \begin{cases} (\nu/\nu_a)^{11/8} & \nu < \nu_a, \\ (\nu/\nu_a)^{-1/2} & \nu_a < \nu < \nu_m, \\ (\nu_m/\nu_a)^{-1/2} (\nu/\nu_m)^{-p/2} & \nu_m < \nu < \nu_{\rm max}. \end{cases} \tag{12}$$

For $r_{\rm cr} \leqslant r \leqslant r_{\rm tr}$,

$$L_\nu^{\rm syn} = L_{\nu_a}^{\rm syn} \begin{cases} (\nu/\nu_a)^{11/8} & \nu < \nu_a, \\ (\nu/\nu_a)^{1/3} & \nu_a < \nu < \nu_c, \\ (\nu_c/\nu_a)^{1/3} (\nu/\nu_c)^{-1/2} & \nu_c < \nu < \nu_m, \\ (\nu_c/\nu_a)^{1/3} (\nu_m/\nu_c)^{-1/2} (\nu/\nu_m)^{-p/2} & \nu_m < \nu < \nu_{\rm max}. \end{cases} \tag{13}$$

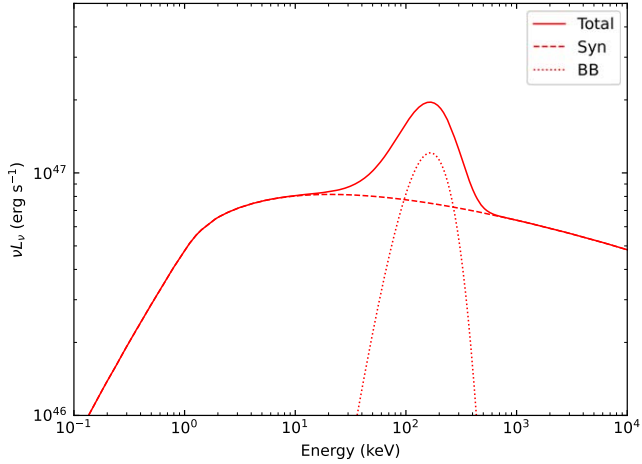

**Figure 1.** Radiation spectrum derived from $B_p = 10^{15}$ G, $P_0 = 1$ ms, $\epsilon_e = 0.1$, $\Gamma_{sat} = 1000$, $p = 2.5$. The solid line represents the total emission, consisting of the nonthermal synchrotron component (dashed line) and the thermal component (dotted line).

For $r_{tr} \leqslant r \leqslant r_{sat}$,

$$L_\nu^{syn} = L_{\nu_a}^{syn} \begin{cases} (\nu/\nu_a)^{11/8} & \nu < \nu_a, \\ (\nu/\nu_a)^{1/3} & \nu_a < \nu < \nu_m, \\ (\nu_m/\nu_a)^{1/3}(\nu/\nu_m)^{-(p-1)/2} & \nu_m < \nu < \nu_c, \\ (\nu_m/\nu_a)^{1/3}(\nu_c/\nu_m)^{-(p-1)/2}(\nu/\nu_c)^{-p/2} & \nu_c < \nu < \nu_{max}. \end{cases} \quad (14)$$

Integrating the above expressions from $r_{ph}$ to $r_{sat}$ gives the nonthermal synchrotron radiation spectrum. As an illustration, in Figure 1 we show the resulting radiation spectrum derived from a set of magnetar and dissipation parameters.

High-energy photons from the striped magnetar wind interact with matter as they diffuse outward through the ejecta. In order to derive the luminosity evolution and energy spectrum of the emergent high-energy emission, we consider the effect of the ejecta material on the absorption and scattering of photons. At ∼0.1 to tens of keV, the X-ray opacity is dominated by photoelectric absorption in matter. From tens of keV to ∼10 MeV, hard X-rays and gamma rays lose most of their energy due to Compton scattering. Gamma rays above ∼10 MeV predominantly lose energy through photon–matter pair production.

In this work, we assume that the ejecta is composed of 20% hydrogen, 10% helium, and 70% oxygen. This assumption mimics the composition of ejecta from electron capture SNe (ECSNe), which are likely to produce a millisecond magnetar as the central engine of AT2018cow, and also incorporates inferences given by the observations (e.g., Margutti et al. 2019). In Figure 2, we plot the frequency-dependent opacities due to various absorption and scattering processes in different neutral media.

The opacity due to bound-free absorption is heavily dependent on the degree of ionization of the ejecta. In general, SN ejecta are highly ionized after a explosion, and as the ejecta expands and cools the ejecta undergoes recombination and eventually becomes dominated by neutral matter. In this case, the opacity due to bound-free absorption is so large that soft X-rays cannot escape. When there is a magnetar as the central engine, the irradiation of energetic photons from the magnetar will result in the ejecta being reionized. As a consequence, the ionization state of the ejecta is controlled by the balance between photoionization and recombination. In the context of SNe, the reionization due to a pulsar wind nebula may make X-rays visible from months to decades after the explosion (e.g., Metzger et al. 2014; Margalit et al. 2018).

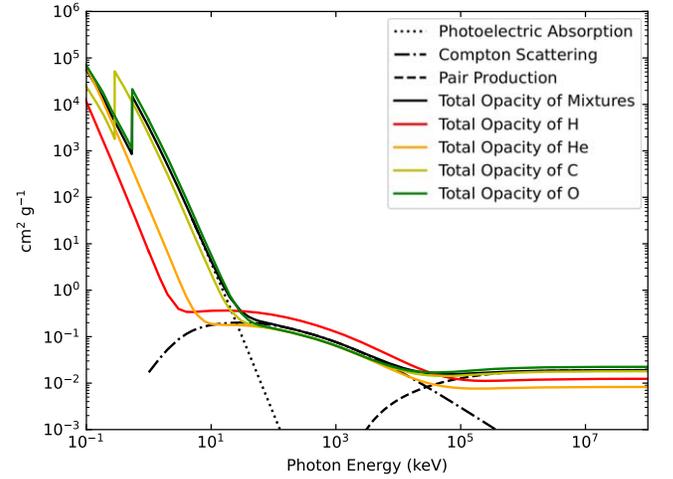

**Figure 2.** Opacity of high-energy photons due to scattering and absorption in H, He, C, O, and mixtures (20%H+10%He+70%O). The dotted line, dashed–dotted line, and dashed line indicate the contributions of photoelectric absorption, Compton scattering, and pair production, respectively. Data for 0.1–1 keV from https://henke.lbl.gov/optical_constants/ and 1 keV–100 GeV from https://www.nist.gov/pml/xcom-photon-cross-sections-database.

The ionization equilibrium at each point in the ejecta is determined by the balance between photoionization and recombination of electrons and ions, which can be written as the ionization equilibrium equation:

$$n^{Z,i} \int \frac{4\pi I_\nu}{h\nu} \sigma_{bf,\nu}^{Z,i} d\nu = \alpha_{rec}^{Z,i} n^{Z,i+1} n_e, \quad (15)$$

where $I_\nu$ is the mean intensity of radiation at each point in the ejecta, $n^{Z,i}$ is the number density for each element $Z$ and ionization state $i$, $\sigma_{bf,\nu}^{Z,i}$ is the photoionization cross section, $\alpha_{rec}^{Z,i}$ is the recombination coefficient, which depends on the ejecta temperature, and $n_e$ is the number density of free electrons. The recombination coefficients for hydrogen and helium are taken from Osterbrock & Ferland (2006), and for oxygen from Nahar (1999). We employed an empirical function $T_{ej} = 10^4(2.38e^{-0.165\left(\frac{t}{day}\right)} + 1.55)$ K to approximate the temperature evolution reported by Margutti et al. (2019). This empirical function consists of two stages: an early exponential decay phase and a later floor temperature phase. The photoionization cross sections are taken from Verner et al. (1996). For simplicity, we assume that the mean intensity is identical at each point in the ejecta, thus giving the following approximation:

$$4\pi I_\nu = \frac{L_\nu}{4\pi R_{ej}^2}, \quad (16)$$

where $R_{ej}$ is the outermost radius of the ejecta. We define the ion fraction

$$f^{Z,i} = \frac{n^{Z,i}}{n^Z} = \frac{4\pi R_{ej}^3 A m_p n^{Z,i}}{3 M_{ej} X_Z}, \quad (17)$$

where $n^Z$ and $X_Z$ are the number density and the mass fraction of species, and $M_{ej}$ are the mass of the ejecta. Thus the ionization equilibrium equation can be rewritten as

$$f^{Z,i} \int \frac{L_\nu}{4\pi R_{ej}^2 h\nu} \sigma_{bf,\nu}^{Z,i} d\nu = n^Z \alpha_{rec}^{Z,i} f^{Z,i+1} \sum i f^{Z,i}. \quad (18)$$

Notice that $\sum f^{Z,i} = 1$, $f^{Z,i}$ can be solved out. Therefore, the total bound-free opacity can be written as

$$\kappa_{bf,\nu} = X_H \frac{1}{m_p} \sum_{i=0}^{1} f^{H,i} \sigma_{bf,\nu}^{H,i} + X_{He} \frac{1}{4m_p} \sum_{i=0}^{2} f^{He,i} \sigma_{bf,\nu}^{He,i}$$
$$+ X_O \frac{1}{16m_p} \sum_{i=0}^{8} f^{O,i} \sigma_{bf,\nu}^{O,i}. \quad (19)$$

Then, to calculate the observed specific luminosity a leakage factor is invoked

$$L_\nu^{obs} = L_\nu e^{-\tau_\nu}, \quad (20)$$

where $\tau_\nu = 3\kappa_\nu M_{ej}/4\pi R_{ej}^2$ is the optical depth of ejecta to high-energy photons, and $\kappa_\nu$ is the opacity of the ejecta.

### 2.2. Emission from the Heated Ejecta

A portion of the high-energy photons produced by the striped magnetar wind is absorbed and scattered by the ejecta, resulting in the ejecta being effectively thermalized. We integrate the trapped specific luminosity from 100 eV to 100 GeV to obtain the input thermal luminosity, which can be written as

$$L_{th,mag} = \int (1 - e^{-\tau_\nu}) L_\nu d\nu. \quad (21)$$

Although observations suggest that the emission from AT2018cow is unlikely to be driven predominantly by the radioactive decay of $^{56}$Ni/Co and limit the mass of $^{56}$Ni to $\lesssim 0.1 M_\odot$ (Margutti et al. 2019), the decay of $^{56}$Co still has the potential to contribute to the late-time optical emission. Here we consider the decay of $^{56}$Ni/Co as an additional energy source to heat the ejecta. The input thermal luminosity for the thermalization of ejecta due to the decay chain $^{56}$Ni → $^{56}$Co → $^{56}$Fe can be written as (Colgate & McKee 1969; Arnett 1980, 1982; Colgate et al. 1980)

$$L_{th,Ni} = m_{Ni}[(\epsilon_{Ni} - \epsilon_{Co})e^{-t/\lambda_{Ni}} + \epsilon_{Co}e^{-t/\lambda_{Co}}](1 - e^{-\tau_\gamma}), \quad (22)$$

where $m_{Ni}$ is the mass of $^{56}$Ni. $\epsilon_{Ni} = 3.9 \times 10^{10}$ erg g$^{-1}$ s$^{-1}$ and $\epsilon_{Co} = 6.78 \times 10^9$ erg g$^{-1}$ s$^{-1}$. $\lambda_{Ni} = 8.8$ days and $\lambda_{Co} = 111.3$ days are the decay time of $^{56}$Ni decays to $^{56}$Co and $^{56}$Co decays to $^{56}$Fe, respectively. $\tau_\gamma$ is the optical depth of the ejecta to gamma rays and can be written as $\tau_\gamma = 3\kappa_\gamma M_{ej}/4\pi R_{ej}^2$. The gray opacity to the $^{56}$Ni/Co cascade decay photons is $\kappa_\gamma = 0.027$ cm$^2$ g$^{-1}$ (Colgate et al. 1980; Swartz et al. 1995).

For homogeneously expanding photosphere and centrally located energy injection, the observed bolometric luminosity can be written in the following form (Arnett 1982; Chatzopoulos et al. 2012):

$$L_{bolo}(t) = \frac{2}{\tau_m} e^{-\frac{t^2}{\tau_m^2}} \int_0^t e^{\frac{t'^2}{\tau_m^2}} \frac{t'}{\tau_m}(L_{th,mag}(t') + L_{th,Ni}(t'))dt', \quad (23)$$

where $\tau_m$ is the effective light-curve timescale and can be written as

$$\tau_m = \left(\frac{10\kappa M_{ej}}{3\beta \nu c}\right)^{1/2}. \quad (24)$$

Here $\kappa$, $v$, and $c$ are the optical opacity, the expansion velocity, and the speed of light, respectively. $\beta \simeq 13.8$ is an approximation for a variety of diffusion mass density profiles. We assume that the opacity is dominated by the Thomson electron scattering opacity $\kappa_{es} = 0.4(\bar{Z}/\bar{A})(x_e/\bar{Z})$ cm$^2$ g$^{-1}$, where $\bar{Z}$ and $\bar{A}$ represent the average atomic number and mass, and $x_e = n_e/n_{nuclei}$ is the ionization fraction. Considering the oxygen-dominated ionized ejecta, we roughly set $\kappa_{es} = 0.2$ cm$^2$ g$^{-1}$ (e.g., Inserra et al. 2013; Nicholl et al. 2017).

### 3. Fitting and Results

Next we perform a model fitting on the multiwavelength data of AT2018cow. The data used for fitting include optical bolometric luminosity derived from blackbody fit to the UV+UBV photometry; soft (0.3–10 keV) and hard X-ray (20–200 keV) luminosity evolution; and a joint energy spectrum of soft and hard X-rays for 7.7 days. The UV/optical data originate from Swift-UVOT (Gehrels et al. 2004; Roming et al. 2005), ANDICAM (DePoy et al. 2003), LRIS (Oke et al. 1995), and DEIMOS (Faber et al. 2003). The soft-to-hard X-ray data originate from Swift-XRT (Gehrels et al. 2004; Burrows et al. 2005), XMM-Newton (Jansen et al. 2001), NuSTAR (Harrison et al. 2013), and INTEGRAL (Lebrun et al. 2003; Ubertini et al. 2003; Winkler et al. 2003). Finally, all data used for this work are taken from Margutti et al. (2019).

We develop a Fortran-based numerical model based on the description in Section 2 and implement the MCMC techniques by using the emcee Python package (Foreman-Mackey et al. 2013). F2PY is employed to provide a connection between Python and Fortran languages. We perform the MCMC with 18 walkers for running at least 100,000 steps, until the step is longer than 50 times the integrated autocorrelation time $\tau_{ac}$, i.e., $N_{step} = \max(10^5, 50\tau_{ac})$ to make sure that the fitting is sufficiently converged (Foreman-Mackey et al. 2013). Once the MCMC is done, the best-fitting values and the 1$\sigma$ uncertainties are computed as the 50th, 16th, and 84th percentiles of the posterior samples.

The free parameters required for the fitting include the total mass of the ejecta ($M_{ej}$); the mass fraction of radioactive $^{56}$Ni ($f_{Ni}$); the expansion velocity of the ejecta ($v_{SN}$); the magnetar surface magnetic field strength at the polar cap region ($B_p$); the magnetar initial spin period ($P_0$); the fraction of dissipated energy per electron ($\epsilon_e$); the bulk Lorentz factor of the wind at the saturation radius ($\Gamma_{sat}$); the energy spectral index of accelerated electrons due to reconnection ($p$); and the time interval between explosion and discovery ($t_{expl}$). Uniform priors on $M_{ej}, f_{Ni}, v_{SN}, \log_{10} B_p, \log_{10} P_0, \epsilon_e, \Gamma_\infty, p, t_{expl}$ are adopted in this paper. In order to explore a parameter space as large as possible, we set a wide enough range for the priors within reasonable limits. The free parameters and priors in our model can be seen in Table 1.

Figures 3 and 4 show the best-fitting multiwavelength light curves and spectrum based on our model. We find that the overall quality of the fitting is good: the theoretical light curves roughly capture the evolutionary behavior of luminosity in AT2018cow; the two-component energy spectrum also matches well the spectral morphology of AT2018cow on day 7.7. To verify whether the

Table 1
Free Parameters, Priors, and Best-fitting Results in Our Model

| Parameter | Prior | Result |
| --- | --- | --- |
| $M_{\rm ej}$ ($M_\odot$) | [0.01, 1] | $0.12^{+0.05}_{-0.03}$ |
| $f_{\rm Ni}$ | [0, 0.5] | $0.50^{+0.00}_{-0.05}$ |
| $v_{\rm SN}$ ($10^9$ cm s$^{-1}$) | [0.1, 10] | $5.11^{+1.14}_{-1.89}$ |
| $\log_{10}[B_p$ (G)$]$ | [10, 16] | $14.38^{+0.36}_{-0.69}$ |
| $\log_{10}[P_0$ (s)$]$ | [−3, −1] | $-2.43^{+0.11}_{-0.20}$ |
| $\epsilon_e$ | [0.01, 0.5] | $0.16^{+0.12}_{-0.08}$ |
| $\Gamma_{\rm sat}$ | [1, 5] | $2.80^{+0.61}_{-0.60}$ |
| $p$ | (2, 3) | $2.41^{+0.10}_{-0.15}$ |
| $t_{\rm expl}$ (days) | [0, 1.31] | $0.47^{+0.56}_{-0.38}$ |
| $\chi^2_{\rm opt+hX}/{\rm dof}$ | ⋯ | 71/46 |
| $\chi^2_{\rm sX}/{\rm dof}$ | ⋯ | 2640/60 |
| $\chi^2_{\rm spec}/{\rm dof}$ | ⋯ | 161/65 |
| $\chi^2_{\rm tot}/{\rm dof}$ | ⋯ | 2872/189 |

**Note.** The uncertainties of the best-fitting parameters are measured as 1$\sigma$ confidence ranges.

late-time soft X-ray spectrum still aligns with our model, we extracted the spectrum from Swift-XRT observations around day 46.7, which is displayed in Figure 4. We obtained the public data set from the High Energy Astrophysics Science Archive Research Center website. The Swift observations were processed using HEASoft 6.32.1. We employed the xrtpipeline and xrtproducts tasks for data product derivation. The selected source region was a circle with a radius of 47″.1 centered on the target, and the background region was an annulus nearby with inner and outer radii of 100″ and 200″, respectively. To enhance the signal-to-noise ratio, we stacked the event files from observational IDs 00010724096-00010724099 and 00010724090-00010724093. The stacked X-ray spectra were derived using xselect. Spectral fitting was conducted in xspec v12.13.1, ensuring a minimum grouping of five photons.

This suggests that, in principle, our model can explain the majority of emission from optical to hard X-rays in AT2018cow. However, our model is unable to account for the persistent soft X-ray flaring with variability timescales of a few days, which may need to be explained by invoking additional physical processes or radiative components. Our model also overestimates the late-time soft X-ray luminosity, which follows a steep power-law decay $\propto t^{-4}$. Table 1 shows the best-fitting chi-square values. Due to the poor fitting quality in the soft X-ray, the total reduced chi-square $\chi^2_{\rm tot}/{\rm dof} \approx 15$ is large (dof stands for degrees of freedom). When counting only optical + hard X-ray light curves and X-ray spectra, we derive reasonable values $\chi^2_{\rm opt+hX}/{\rm dof} \approx 2$ and $\chi^2_{\rm spec}/{\rm dof} \approx 2$, respectively. When only soft X-ray light curves are considered, we derive a large value of $\chi^2_{\rm sX}/{\rm dof} \approx 44$.

Table 1 shows the best-fitting parameters and their corresponding 1$\sigma$ uncertainties. We find that in order to explain the very short rise time to peak of the thermal optical emission, a low ejecta mass $M_{\rm ej} \approx 0.1\,M_\odot$ and high ejecta velocity $v_{\rm SN} \approx 0.17c$ are required. A millisecond magnetar with $P_0 \approx 3.7$ ms and $B_p \approx 2.4 \times 10^{14}$ G is needed to serve as the central engine of AT2018cow. The observed optical, soft X-ray, and hard X-ray emission is mainly driven by internal gradual magnetic dissipation in the striped magnetar wind. During internal dissipation, electrons are accelerated due to magnetic reconnection. A fraction $\epsilon_e \approx 0.16$ of the dissipated energy per electron is used to accelerate and result in a power-law distribution of electrons with an index $p \approx 2.4$. Meanwhile, the magnetar wind obtains a bulk kinetic energy with a terminal Lorentz factor $\Gamma_{\rm sat} \approx 630$. In our model, the radioactive decay of $^{56}$Ni/Co as an additional energy source provides a partial contribution to the optical emission of AT2018cow. However, the model-predicted late-time luminosity is still slightly lower than the observed bolometric luminosity, even though the mass fraction $f_{\rm Ni} \approx 0.5$ (corresponding to the $^{56}$Ni mass of $M_{\rm Ni} \approx 0.05$) reaches the upper limit of the prior we have set. Last, to visualize how the energy of the magnetar is allocated to the different bands at different times, we show the evolution of the value of optical depth over time in Figure 5. After a few days, the ejecta turned transparent to energies ranging from 0.1 to 200 keV.

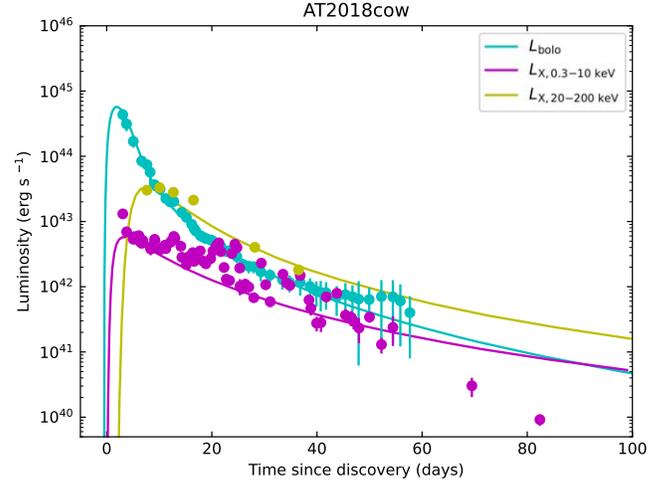

**Figure 3.** Model fit to the luminosity evolution of the different emission components of AT2018cow: optical bolometric luminosity derived from a blackbody fit to the UV/optical photometry (cyan), soft X-ray (0.3–10 keV; purple), hard X-ray (20–200 keV; yellow).

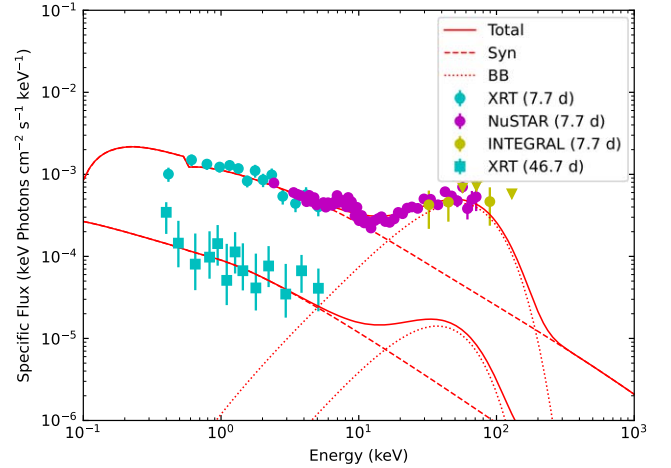

**Figure 4.** Model fit to the broadband X-ray spectrum of AT2018cow at day 7.7, and comparison of soft X-ray spectrum and model on day 46.7. Data from Swift-XRT, NuSTAR, and INTEGRAL are shown in cyan, purple, and yellow, respectively.

## 4. Discussion and Conclusions

### 4.1. Soft X-Ray Flares

The largest shortcoming of our model is that it poorly reproduces the soft X-ray luminosity evolution. In the soft

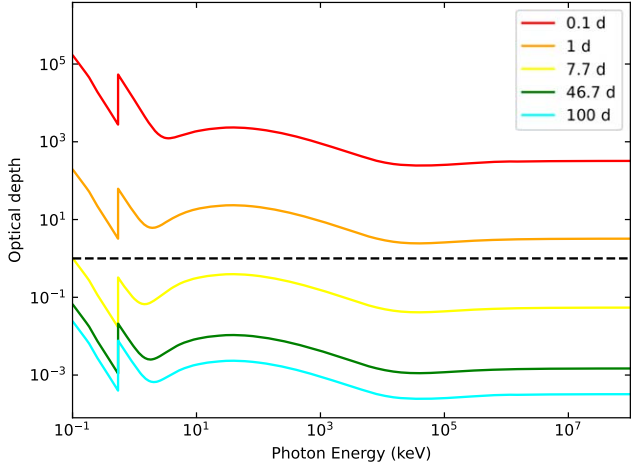
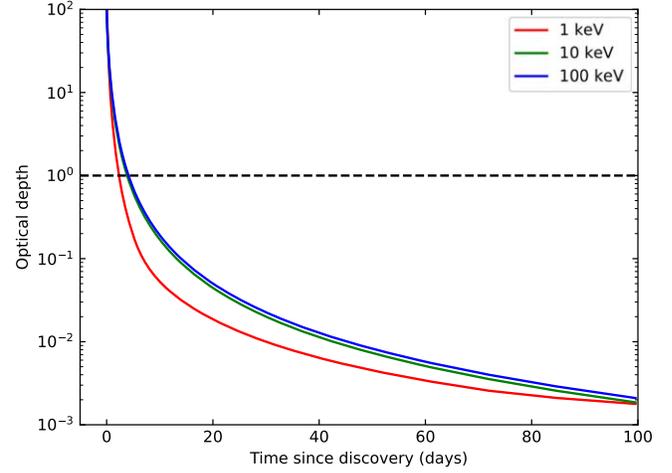

**Figure 5.** Left panel: the optical depth as a function of photon energy at specific times: 0.1, 1, 7.7, 46.7, and 100 days. Right panel: the evolution of optical depth over time for photons at 1, 10, and 100 keV.

X-ray band, there are persistent flares with a variability timescale of a few days, superimposed on a secular power-law decay. In addition, these flares appear to be quasiperiodic, with the most significant period being about 4 days with a $3\sigma$ confidence level (Kuin et al. 2019; Margutti et al. 2019). These quasiperiodic flares may be related to magnetar free precession. There are several studies suggesting that newborn millisecond magnetar precession could explain AT2018cow-like soft X-ray flares in some GRB afterglows (e.g., Suvorov & Kokkotas 2020, 2021; Zou et al. 2021; Zou & Liang 2022). In our model, free precession may lead to variability in the observed X-ray luminosity through the following three reasons: fluctuations in spin-down luminosity due to the evolution of the inclination angle between the rotation and the magnetic axes; variation of the dissipation region; and Doppler boosting/deboosting due to the relativistic bulk motion of the wind. In principle, the variability timescale can be connected to the precession period $P_{\rm prec} \sim P/\epsilon$. If we set $P_{\rm prec} \sim 4$ days, and the spin period $P \sim 10$ ms at $\sim 10$–60 days, then the ellipticity $\epsilon \sim 3 \times 10^{-8}$. This order of magnitude of nonsphericity may be caused by the internal magnetic field of the magnetar, which can be written as (e.g., Zanazzi & Lai 2015)

$$\epsilon_{\rm mag} = \beta \frac{R_{\rm NS}^4 B_*^2}{GM_{\rm NS}^2}$$

$$= 2 \times 10^{-8} \beta \left(\frac{B_*}{10^{14}\,{\rm G}}\right)^2 \left(\frac{R_{\rm NS}}{10^6\,{\rm cm}}\right)^4 \left(\frac{M_{\rm NS}}{1.4\,M_\odot}\right)^{-2}, \quad (25)$$

where $B_*$ is the internal magnetic field strength, and $\beta$ is a dimensionless constant that depends on the geometry of the internal magnetic field, typically approaching unity for a dipole or toroidal topology (Mastrano et al. 2013).

While our model considers the precession effect in explaining the soft X-ray data, these factors might lead to the absence of fluctuation effects in hard X-ray and optical data. The absence of fluctuations in the hard X-ray data could be attributed to the relatively fewer data points in the hard X-ray range. The absence of fluctuations in optical data might be due to the weakening of this effect through the absorption and reprocessing of high-energy photons by the ejecta. Additionally, the optical emission is also influenced by the decay of $^{56}$Ni and $^{56}$Co, which could further diminish the fluctuation effect.

### 4.2. Late-time Steep Decay

The late-time steep decay recalls similar behavior in some GRBs. The plateau followed by a steep decay in some GRB X-ray afterglows cannot be explained in the framework of external shock models and usually have to invoke the internal dissipation process of the magnetar wind, while the steep decay may be related to an abrupt cutoff of the center engine (such as the magnetar collapsing into a BH; e.g., Troja et al. 2007; Rowlinson et al. 2010, 2013; Lü & Zhang 2014; Lü et al. 2015; Chen et al. 2017; Lan et al. 2020). For the case of AT2018cow, the power off of the center engine immediately ceases the production of high-energy photons from the magnetic dissipation. However, it may take some time for this information to propagate to the surface of the ejecta. This would correspond to a steep decay of the observed X-ray luminosity as the remaining photons in the ejecta from the outer to the inner continue to propagate out.

If the central engine of AT2018cow is a massive NS with a mass exceeding the maximum mass of a nonrotating NS ($M_{\rm TOV}$), then it can only survive for a limited time $T_c$. For a given equation of state (EOS), the maximum gravitational mass ($M_{\rm max}$) can be expressed as (Lyford et al. 2003; Lasky et al. 2014)

$$M_{\rm max} = M_{\rm TOV}(1 + \hat{\alpha} P^{\hat{\beta}}), \quad (26)$$

where $\hat{\alpha}$ and $\hat{\beta}$ are parameters related to the EOS. $M_{\rm max}$ decreases as the NS spin down. As $M_{\rm max}$ approaches the NS mass, the centrifugal force will no longer be able to sustain gravity and the NS will collapse into a BH. Substituting the magnetic dipole radiation dominated spin evolution $P(t) = P_0(1 + t/\tau_{\rm sd})^{1/2}$ into the above equation, one can derive the collapse timescale

$$T_c = \frac{3c^3 I}{4\pi^2 B_p^2 R^6}\left[\left(\frac{M - M_{\rm TOV}}{\hat{\alpha} M_{\rm TOV}}\right)^{2/\hat{\beta}} - P_0^2\right]. \quad (27)$$

In order to match the observations, the collapse timescale is assumed to be $T_c \approx 50$ days. We choose a set of parameters corresponding to one possible EOS (GM1; Ai et al. 2018): $M_{\rm TOV} = 2.37\,M_\odot$; $R = 12.05$ km; $I = 3.33 \times 10^{45}$ g cm$^2$; $\hat{\alpha} = 1.58 \times 10^{-10}$ s$^{-\hat{\beta}}$; $\hat{\beta} = -2.84$. For the best-fitting magnetar parameters we derived, the corresponding NS mass $M_{NS} \approx 2.3701\, M_\odot$. The derived NS mass is very close to the $M_{TOV}$ since the spin-down timescale $\tau_{sd} \approx 2$ days is much smaller than the collapse timescale. Thus the engine shutdown due to the collapse of the NS into a BH is barely possible.

The rapid decline in the soft X-ray late-time might be attributed to other factors. For instance, an increase in the recombination rate or a decrease in the ionization rate of the ejecta could lead to a lower ionization fraction, thereby increasing the opacity in the soft X-ray late-time and contributing to its rapid decline. Our model, due to its simplifications and limited capability to replicate the actual scenario, currently cannot reproduce this phenomenon. Another possible explanation could be the decay of the magnetic field due to material accreting onto the magnetar's surface (e.g., Taam & van den Heuvel 1986; Shibazaki et al. 1989; Fu & Li 2013).

### 4.3. Possible Progenitor Model

Considering the presence of hydrogen inferred from observations and the low ejecta mass derived from our model, it is possible that AT2018cow is associated with an ECSN giving birth to a millisecond magnetar. ECSNe that typically arise in stars with initial masses of $\sim 8\text{–}10\, M_\odot$ have ejecta masses $M_{ej} \lesssim 1\, M_\odot$ and kinetic energies $E_{kin} \sim 10^{50}$ erg. Our model gives the ejecta kinetic energy $\approx 7 \times 10^{50}$ erg, which suggests that part of the kinetic energy may be provided by the center engine. Furthermore, observations suggest that the ejecta of AT2018cow is asymmetric (Maund et al. 2023), with fast, low-density ejecta along the polar direction and slow, dense ejecta in the equatorial plane (Margutti et al. 2019). Our derived ejecta parameters should correspond to the fast component, and the total ejecta mass would be larger when the slow ejecta are taken into account. Both the asymmetric ejecta of AT2018cow and the presence of the millisecond magnetar as its central engine imply that its progenitor may have been a rapidly rotating star, which also explains why such an event is so rare.

To summarize, our study suggests that a millisecond magnetar could serve as the central engine of AT2018cow, highlighting the multiwavelength emission phenomena observed. While our model could not perfectly match the multiwavelength data of AT2018cow, it effectively captured the general trends in its luminosity evolution and energy spectral features. The findings underscore the potential role of the magnetar as the central engine, with the internal dissipation of the magnetar wind playing a crucial role in driving the multiwavelength emission observed in AT2018cow. It may be reasonable not considering the magnetar wind nebula emission in our model, since the magnetic dissipation in the striped magnetar wind is effective. The result is that the magnetization in the nebula may be very low, which is necessary to explain the the observed late-time steepening of superluminous SN optical light curves (Vurm & Metzger 2021). It should also be emphasized that AT2018cow has many other peculiar observational properties that are not discussed in the present work, such as high-amplitude quasiperiodic oscillation of AT2018cow's soft X-rays with a frequency of 224 Hz (Pasham et al. 2021); possible 250 s X-ray quasiperiodicity (Zhang et al. 2022); luminous UV emission detected 2–4 yr after the explosion (Sun et al. 2022; Chen et al. 2023); and persistent X-ray emission at the location of AT2018cow $\sim 3.7$ yr after discovery (Migliori et al. 2023). It is a great challenge to explain all these peculiar observational properties of AT2018cow based on the millisecond magnetar as the central engine and there is still a lot of work to be done. In the future, as more AT2018cow-like events are discovered, it will help to understand the central engine of these FBOTs as well as their progenitor systems.

### Acknowledgments

We thank the referee for the suggestions that helped us improve this manuscript significantly. This work is supported by National SKA Program of China (grant No. 2020SKA0120300) and National Natural Science Foundation of China (grant No. 11833003). L.L. acknowledge support from the National Natural Science Foundation of China (grant No. 12303050), China Postdoctoral Science Foundation (grant No. 2023M743397), and the Fundamental Research Funds for the Central Universities. S.Q.Z. is supported by the National Natural Science Foundation of China (grant No. 12247144) and China Postdoctoral Science Foundation (grant Nos. 2021TQ0325 and 2022M723060). D.X. is supported by the National Natural Science Foundation of China (grant No. 12373052). S.F.H. is supported by the Anhui Provincial Natural Science Foundation (2308085QA32). Z.F.S. is supported by the National Natural Science Foundation of China (grant No. 12103048) and the National Key R&D Program of China (No. 2023YFA1608100).

*Software:* emcee (Foreman-Mackey et al. 2013).

### ORCID iDs

Long Li https://orcid.org/0000-0002-8391-5980
Shu-Qing Zhong https://orcid.org/0000-0002-1766-6947
Di Xiao https://orcid.org/0000-0002-4304-2759
Zi-Gao Dai https://orcid.org/0000-0002-7835-8585
Shi-Feng Huang https://orcid.org/0000-0001-7689-6382
Zhen-Feng Sheng https://orcid.org/0000-0001-6938-8670

### References

Ai, S., Gao, H., Dai, Z.-G., et al. 2018, ApJ, 860, 57
Antoni, A., & Quataert, E. 2022, MNRAS, 511, 176
Arcavi, I., Wolf, W. M., Howell, D. A., et al. 2016, ApJ, 819, 35
Arnett, W. D. 1980, ApJ, 237, 541
Arnett, W. D. 1982, ApJ, 253, 785
Beniamini, P., & Giannios, D. 2017, MNRAS, 468, 3202
Beniamini, P., & Piran, T. 2014, MNRAS, 445, 3892
Burrows, D. N., Hill, J. E., Nousek, J. A., et al. 2005, SSRv, 120, 165
Chatzopoulos, E., Wheeler, J. C., & Vinko, J. 2012, ApJ, 746, 121
Chen, W., Xie, W., Lei, W.-H., et al. 2017, ApJ, 849, 119
Chen, Y., Drout, M. R., Piro, A. L., et al. 2023, ApJ, 955, 43
Colgate, S. A., & McKee, C. 1969, ApJ, 157, 623
Colgate, S. A., Petschek, A. G., & Kriese, J. T. 1980, ApJL, 237, L81
Coroniti, F. V. 1990, ApJ, 349, 538
Dai, Z. G. 2004, ApJ, 606, 1000
Dai, Z. G., & Lu, T. 1998a, PhRvL, 81, 4301
Dai, Z. G., & Lu, T. 1998b, A&A, 333, L87
DePoy, D. L., Atwood, B., Belville, S. R., et al. 2003, Proc. SPIE, 4841, 827
Drenkhahn, G. 2002, A&A, 387, 714
Drenkhahn, G., & Spruit, H. C. 2002, A&A, 391, 1141
Drout, M. R., Chornock, R., Soderberg, A. M., et al. 2014, ApJ, 794, 23
Faber, S. M., Phillips, A. C., Kibrick, R. I., et al. 2003, Proc. SPIE, 4841, 1657
Fang, K., Metzger, B. D., Murase, K., Bartos, I., & Kotera, K. 2019, ApJ, 878, 34
Foreman-Mackey, D., Hogg, D. W., Lang, D., & Goodman, J. 2013, PASP, 125, 306
Fox, O. D., & Smith, N. 2019, MNRAS, 488, 3772
Fu, L., & Li, X.-D. 2013, ApJ, 775, 124

Gehrels, N., Chincarini, G., Giommi, P., et al. 2004, ApJ, 611, 1005
Geng, J.-J., Dai, Z.-G., Huang, Y.-F., et al. 2018, ApJL, 856, L33
Giannios, D. 2006, A&A, 457, 763
Giannios, D. 2008, A&A, 480, 305
Giannios, D. 2010, MNRAS, 408, L46
Giannios, D. 2012, MNRAS, 422, 3092
Giannios, D., & Spruit, H. C. 2005, A&A, 430, 1
Gottlieb, O., Tchekhovskoy, A., & Margutti, R. 2022, MNRAS, 513, 3810
Granot, J., & Sari, R. 2002, ApJ, 568, 820
Guo, F., Liu, Y.-H., Daughton, W., & Li, H. 2015, ApJ, 806, 167
Harrison, F. A., Craig, W. W., Christensen, F. E., et al. 2013, ApJ, 770, 103
Ho, A. Y. Q., Perley, D. A., Gal-Yam, A., et al. 2023, ApJ, 949, 120
Ho, A. Y. Q., Phinney, E. S., Ravi, V., et al. 2019, ApJ, 871, 73
Inserra, C. 2019, NatAs, 3, 697
Inserra, C., Smartt, S. J., Jerkstrand, A., et al. 2013, ApJ, 770, 128
Jansen, F., Lumb, D., Altieri, B., et al. 2001, A&A, 365, L1
Kagan, D., Sironi, L., Cerutti, B., & Giannios, D. 2015, SSRv, 191, 545
Kotera, K., Phinney, E. S., & Olinto, A. V. 2013, MNRAS, 432, 3228
Kremer, K., Lu, W., Piro, A. L., et al. 2021, ApJ, 911, 104
Kuin, N. P. M., Wu, K., Oates, S., et al. 2019, MNRAS, 487, 2505
Lan, L., Lü, H.-J., Rice, J., & Liang, E.-W. 2020, ApJ, 890, 99
Lander, S. K., & Jones, D. I. 2020, MNRAS, 494, 4838
Lasky, P. D., Haskell, B., Ravi, V., Howell, E. J., & Coward, D. M. 2014, PhRvD, 89, 047302
Lebrun, F., Leray, J. P., Lavocat, P., et al. 2003, A&A, 411, L141
Leung, S.-C., Blinnikov, S., Nomoto, K., et al. 2020, ApJ, 903, 66
Li, L., & Dai, Z.-G. 2021, ApJ, 918, 52
Liu, Y.-H., Guo, F., Daughton, W., Li, H., & Hesse, M. 2015, PhRvL, 114, 095002
Lü, H.-J., & Zhang, B. 2014, ApJ, 785, 74
Lü, H.-J., Zhang, B., Lei, W.-H., Li, Y., & Lasky, P. D. 2015, ApJ, 805, 89
Lyford, N. D., Baumgarte, T. W., & Shapiro, S. L. 2003, ApJ, 583, 410
Lyubarsky, Y. E. 2005, MNRAS, 358, 113
Lyutikov, M. 2022, MNRAS, 515, 2293
Lyutikov, M., & Toonen, S. 2019, MNRAS, 487, 5618
Margalit, B., Metzger, B. D., Berger, E., et al. 2018, MNRAS, 481, 2407
Margutti, R., Metzger, B. D., Chornock, R., et al. 2019, ApJ, 872, 18
Mastrano, A., Lasky, P. D., & Melatos, A. 2013, MNRAS, 434, 1658
Maund, J. R., Höflich, P. A., Steele, I. A., et al. 2023, MNRAS, 521, 3323
Metzger, B. D. 2022, ApJ, 932, 84
Metzger, B. D., Giannios, D., Thompson, T. A., Bucciantini, N., & Quataert, E. 2011, MNRAS, 413, 2031
Metzger, B. D., & Perley, D. A. 2023, ApJ, 944, 74
Metzger, B. D., Vurm, I., Hascoët, R., & Beloborodov, A. M. 2014, MNRAS, 437, 703
Migliori, G., Margutti, R., Metzger, B. D., et al. 2023, arXiv:2309.15678
Mohan, P., An, T., & Yang, J. 2020, ApJL, 888, L24
Murase, K., Kashiyama, K., Kiuchi, K., & Bartos, I. 2015, ApJ, 805, 82
Nahar, S. N. 1999, ApJS, 120, 131
Nayana, A. J., & Chandra, P. 2021, ApJL, 912, L9
Nicholl, M., Guillochon, J., & Berger, E. 2017, ApJ, 850, 55
Oke, J. B., Cohen, J. G., Carr, M., et al. 1995, PASP, 107, 375
Osterbrock, D. E., & Ferland, G. J. 2006, Astrophysics of Gaseous Nebulae and Active Galactic Nuclei (Sausalito, CA: Univ. Science Books)
Pasham, D. R., Ho, W. C. G., Alston, W., et al. 2021, NatAs, 6, 249
Perley, D. A., Mazzali, P. A., Yan, L., et al. 2019, MNRAS, 484, 1031
Piro, A. L., & Lu, W. 2020, ApJ, 894, 2
Prentice, S. J., Maguire, K., Smartt, S. J., et al. 2018, ApJL, 865, L3
Pursiainen, M., Childress, M., Smith, M., et al. 2018, MNRAS, 481, 894
Quataert, E., Lecoanet, D., & Coughlin, E. R. 2019, MNRAS, 485, L83
Rivera Sandoval, L. E., Maccarone, T. J., Corsi, A., et al. 2018, MNRAS, 480, L146
Roming, P. W. A., Kennedy, T. E., Mason, K. O., et al. 2005, SSRv, 120, 95
Rowlinson, A., O'Brien, P. T., Metzger, B. D., Tanvir, N. R., & Levan, A. J. 2013, MNRAS, 430, 1061
Rowlinson, A., O'Brien, P. T., Tanvir, N. R., et al. 2010, MNRAS, 409, 531
Sari, R., Piran, T., & Narayan, R. 1998, ApJL, 497, L17
Schrøder, S. L., MacLeod, M., Loeb, A., Vigna-Gómez, A., & Mandel, I. 2020, ApJ, 892, 13
Shapiro, S. L., & Teukolsky, S. A. 1983, Black Holes, White Dwarfs and Neutron Stars. The Physics of Compact Objects (New York: Wiley)
Shibazaki, N., Murakami, T., Shaham, J., & Nomoto, K. 1989, Natur, 342, 656
Sironi, L., Petropoulou, M., & Giannios, D. 2015, MNRAS, 450, 183
Sironi, L., & Spitkovsky, A. 2014, ApJL, 783, L21
Soker, N. 2022, RAA, 22, 055010
Soker, N., Grichener, A., & Gilkis, A. 2019, MNRAS, 484, 4972
Spruit, H. C., Daigne, F., & Drenkhahn, G. 2001, A&A, 369, 694
Sun, N.-C., Maund, J. R., Crowther, P. A., & Liu, L.-D. 2022, MNRAS, 512, L66
Suvorov, A. G., & Kokkotas, K. D. 2020, ApJL, 892, L34
Suvorov, A. G., & Kokkotas, K. D. 2021, MNRAS, 502, 2482
Swartz, D. A., Sutherland, P. G., & Harkness, R. P. 1995, ApJ, 446, 766
Taam, R. E., & van den Heuvel, E. P. J. 1986, ApJ, 305, 235
Tanaka, M., Tominaga, N., Morokuma, T., et al. 2016, ApJ, 819, 5
Troja, E., Cusumano, G., O'Brien, P. T., et al. 2007, ApJ, 665, 599
Ubertini, P., Lebrun, F., Di Cocco, G., et al. 2003, A&A, 411, L131
Uno, K., & Maeda, K. 2020, ApJ, 897, 156
Verner, D. A., Ferland, G. J., Korista, K. T., & Yakovlev, D. G. 1996, ApJ, 465, 487
Vurm, I., & Metzger, B. D. 2021, ApJ, 917, 77
Werner, G. R., Uzdensky, D. A., Cerutti, B., Nalewajko, K., & Begelman, M. C. 2016, ApJL, 816, L8
Winkler, C., Courvoisier, T. J. L., Di Cocco, G., et al. 2003, A&A, 411, L1
Xiao, D., & Dai, Z.-G. 2017, ApJ, 846, 130
Xiao, D., & Dai, Z.-G. 2019, ApJ, 878, 62
Xiao, D., Peng, Z.-k., Zhang, B.-B., & Dai, Z.-G. 2018, ApJ, 867, 52
Xiao, D., Zhang, B.-B., & Dai, Z.-G. 2019, ApJL, 879, L7
Yu, Y.-W., Chen, A., & Li, X.-D. 2019, ApJL, 877, L21
Yu, Y. W., & Dai, Z. G. 2007, A&A, 470, 119
Zanazzi, J. J., & Lai, D. 2015, MNRAS, 451, 695
Zhang, B., & Mészáros, P. 2001, ApJL, 552, L35
Zhang, W., Shu, X., Chen, J.-H., et al. 2022, RAA, 22, 125016
Zou, L., & Liang, E.-W. 2022, MNRAS, 513, L89
Zou, L., Zheng, T.-C., Yang, X., et al. 2021, ApJL, 921, L1